\documentclass{sig-alternate} 

\usepackage{listings}
\usepackage{array}
\usepackage{amsmath,amssymb}
\usepackage{semantic}
\usepackage{alltt}
\usepackage{tikz}
\usepackage{multirow}
\newcommand{\url}[1]{\texttt{#1}}
\usepackage{stmaryrd}

\newcommand{\optimized}{\emph{OPT}}
\newcommand{\fmoods}{\emph{BASIC}}

\CopyrightYear{2008}
\crdata{978-1-60558-382-2/08/11}
\conferenceinfo{PASTE '08,}{Atlanta, Georgia USA}
\mathlig{|->}{\mapsto}
\newcommand{\leino}{F{\"a}hndrich and Leino}

\newcommand{\ab}[1]{#1^\sharp}
\newcommand{\ia}[1]{ |[#1 |]}

\newcommand{\safe}{\text{safe}}

\newcommand{\UnDef}{\text{UnDef}}
\newcommand{\Def}{\text{Def}}
\newcommand{\NotNull}{\text{NonNull}}
\newcommand{\Null}{\text{Nullable}}
\newcommand{\NullInit}{\text{NullableInit}}
\newcommand{\NULL}{\mathbf{null}}
\newcommand{\isdef}{\mathbf{def}}
\newcommand{\isundef}{\mathbf{undef}}

\newcommand{\fields}{\text{fields}}
\renewcommand{\H}{\ab{H}}
\newcommand{\dom}{\text{dom}}
\newcommand{\T}{\ab{T}}
\newcommand{\LV}{\ab{L}}
\newcommand{\MV}{\ab{M}}

\newcommand{\Raw}{\text{Raw}}
\newcommand{\raw}{\text{Raw}}
\newcommand{\rawtop}{\text{Raw$^{-}$}}

\newcommand{\Val}{\text{Val}}
\newcommand{\TVal}{\ab{\text{TVal}}}
\newcommand{\Loc}{\text{Loc}}
\newcommand{\Class}{\text{Class}}
\newcommand{\class}{\text{class}}

\newcommand{\name}{\text{name}}
\newcommand{\Method}{\text{Method}}

\newcommand{\LocalVar}{\text{LocalVar}}
\newcommand{\Heap}{\text{Heap}}
\newcommand{\State}{\text{State}}
\newcommand{\Object}{\text{Object}}

\newcommand{\Error}{\Omega}
\newcommand{\minit}{\textbf{init}}
\newcommand{\this}{\textbf{this}}
\newcommand{\F}{\mathbb{F}}
\newcommand{\C}{\mathbb{C}}
\newcommand{\V}{\mathbb{V}}
\newcommand{\M}{\mathbb{M}}
\newcommand{\PC}{\mathbb{N}}

\newcommand{\struct}[1]{\{\!| #1 |\!\}}

\newcommand{\IsDef}{\text{IsDef}}
\newcommand{\bnew}{\ensuremath{\mathbf{new}}}
\newcommand{\bif}{\mathbf{if}}
\newcommand{\bret}{\mathbf{return}}

\newcommand{\meth}{\text{m}}
\newcommand{\main}{\textbf{main}}

\newcommand{\semantics}[1]{\llbracket{}#1\rrbracket}

\newcommand{\pto}{\rightharpoonup}

\begin{document}

\title{A Non-Null Annotation Inferencer for Java Bytecode
  \titlenote{This work was supported in part by the R{\'e}gion
    Bretagne}}

\numberofauthors{1}
\author{
  \alignauthor
  Laurent Hubert\\
  \affaddr{CNRS/IRISA}\\
  \affaddr{Campus Beaulieu}\\
  \affaddr{35042 Rennes, France}\\
  \email{laurent.hubert@irisa.fr}\\
}

\maketitle{}

\begin{abstract}
  We present a non-null annotations inferencer for the Java bytecode
  language.
  We previously proposed an analysis to infer non-null annotations and
  proved it soundness and completeness with respect to a state of the
  art type system.  This paper proposes extensions to our former
  analysis in order to deal with the Java bytecode language.  We have
  implemented both analyses and compared their behaviour on several
  benchmarks.  The results show a substantial improvement in the
  precision and, despite being a whole-program analysis, production
  applications can be analyzed within minutes.
\end{abstract}



\section{Introduction}
\label{sec:introduction}

A common source of exceptional program behaviour is the dereferencing
of null references (also called null pointers), resulting in
segmentation faults in C or null pointer exceptions in Java.  Even if
such exceptions are caught, exception handlers represents additional
branches which can make verification more difficult (bigger
verification conditions, implicit flow in information flow
verification, etc.) and disable some optimizations.  Furthermore, the
Java virtual machine is obliged to perform run-time checks for
non-nullness of references when executing a number of its bytecode
instructions, thereby incurring a performance penalty.

As a solution, \leino{} proposed a type system~\cite{FahndrichL03} to
partition references between those which may contain the null constant
and those which may not.  The user had to annotate the program with
non-null types which was not practical in the case of legacy code.
We proposed a formal definition of this type
system~\cite{hubert08-1:nonnull_annotations_inference}, proved its
soundness and provided a whole-program dataflow analysis to infer
those annotations.  


The contribution we present is the tool NIT\footnote{The tool has been
  released under the GNU General Public License and is available at
  \url{http://nit.gforge.inria.fr}} (Nullability Inference Tool), an
implementation resulting from our provably sound analysis.
This analysis had been designed on a relatively high level language,
well suited for the definition of the analysis and the proofs but too
far from the target language of the implementation: the Java bytecode.
After recalling the sound analysis this paper is based on, we describe
the improvements we have brought to the former analysis.  We then
present the implementations of both analyses and compare the results
of the two implementations on practical benchmarks.


\section{A Sound Inference Analysis}

%
%

\subsection{Non-Null Annotations}
\label{sec:nonnull-annotations}

The main difficulty in building a precise non-null type system for
Java is that all objects have their reference fields set to null at
the beginning of their lifetime.  Explicit initialization of fields
usually occurs during the execution of the constructor, potentially in
a method called from the constructor, but it is not mandatory and a
field may be read before it is explicitly initialized ---~in which
case it holds the null constant.  We consider a field that has not
been explicitly initialized during the execution of the constructor to
be implicitly initialized to null at the end of the constructors.

\begin{figure*}
  \centering
  \begin{tabular}{@{}ccc@{}}
\begin{minipage}{.32\linewidth}
\tt
class C $\lbrace$\\
\hspace*{1.4ex}@Nullable Object f;\\
\hspace*{1.4ex}C()$\lbrace$\\
\hspace*{2.8ex}this.f = new Object();\\
\hspace*{1.4ex}$\rbrace$\\
\hspace*{1.2ex}@Nullable Object m(@NonNull C x)$\lbrace$\\
\hspace*{2.8ex}return x.f;\\
\hspace*{1.4ex}$\rbrace$\\
$\rbrace$\\
\end{minipage}&
\begin{minipage}{.31\linewidth}
\tt
class C $\lbrace$\\
\hspace*{1.4ex}@NonNull Object f;\\
\hspace*{1.4ex}C()$\lbrace$\\
\hspace*{2.8ex}m(this);\\
\hspace*{2.8ex}this.f = new Object();\\
\hspace*{1.4ex}$\rbrace$\\
\hspace*{1.4ex}@NonNull Object m(@NonNull C x)$\lbrace$\\
\hspace*{2.8ex}return x.f;\\
\hspace*{1.4ex}$\rbrace$\\
$\rbrace$\\
\end{minipage}&
\begin{minipage}{.31\linewidth}
\tt
class C $\lbrace$\\
\hspace*{1.4ex}@NonNull Object f;\\
\hspace*{1.4ex}C()$\lbrace$\\
\hspace*{2.8ex}m(this);\\
\hspace*{2.8ex}this.f = new Object();\\
\hspace*{1.4ex}$\rbrace$\\
\hspace*{1.4ex}@Nullable Object m(@Raw C x)$\lbrace$\\
\hspace*{2.8ex}return x.f;\\
\hspace*{1.4ex}$\rbrace$\\
$\rbrace$\\
\end{minipage}\\
(a) Too Weak Annotations&
(b) Motivation for \texttt{@Raw} Annotations&
(c) \texttt{@Raw} Annotations
  \end{tabular}
  \caption{Motivating Examples}
  \label{fig:examples}
\end{figure*}

\begin{figure}
  \centering
  \begin{tikzpicture}[scale=0.7]
    \draw (1,3) rectangle (6,4);
    \draw (6,4) -- (8,4);
    \draw (6,3) -- (8,3);
    \draw[style=dashed] (8,4) -- (10,4);
    \draw[style=dashed] (8,3) -- (10,3);
    \draw[->] (1,2) node[anchor=north,draw] {
      \begin{tabular}{@{}l@{}}\verb|class:C|\\\verb|C.f:null|\end{tabular}} -- (1,3);
    \draw[->] (4,2) node[anchor=north,draw] {
      \begin{tabular}{@{}l@{}}\verb|class:C|\\\verb|C.f:|$v$\end{tabular}} -- (4,3);
    \draw[->] (7,2) node[anchor=north,draw] {
      \begin{tabular}{@{}l@{}}\verb|class:C|\\\verb|C.f:|$w$\end{tabular}} -- (7,3);
    \draw (3.5,3) node[above=4.5pt] {\emph{initialization}};
    \draw (8,3) node[above=3pt] {\emph{rest of life}};
  \end{tikzpicture}
  \caption{Lifetime of an Object}
  \label{fig:life}
\end{figure}
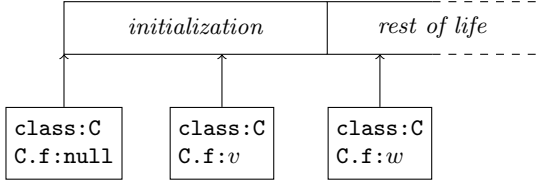

Figures~\ref{fig:examples}(a) shows a class \texttt{C} while
Fig.~\ref{fig:life} shows a model of the lifetime of an instance of
class \texttt{C}.  Assume no other method writes to \texttt{C.f}.
The first part of the object's lifetime is the execution of the
constructor, which is mandatory and occurs to all objects.  The field
\texttt{f} is always explicitly initialized in the constructor and
never written elsewhere, so any read of field \texttt{f} will yield a
non-null reference.  Despite that, if an annotation must be given for
this field valid for the whole lifetime of the object, it will have to
represent the non-null reference put in the field by the constructor,
but also the default null constant present at the beginning of the
object's lifetime.  Such an annotation is basically \texttt{@Nullable}
where we would have clearly preferred a more precise information, such
as \texttt{@NonNull}.
The solution is to consider that annotations on fields are only valid
after the end of the constructor, during the rest of the object's
lifetime.  The \texttt{@Nullable} annotations can now be replaced with
\texttt{@NonNull} annotations.

Figure~\ref{fig:examples}(b) shows the same class where
\texttt{@Nullable} annotations have been replaced and a call to the
method \texttt{m} has been added to the constructor.  The method
\texttt{m} simply reads and returns the value of \texttt{f}.
Although this method is not a constructor, the object \texttt{x} may
still be in its construction phase, \emph{i.e.} the constructor of the
object from which the field is read may not have been fully executed,
and the value actually read may be null in contradiction with the
field annotation.
In fact, for each variable, we need to know whether the reference may
point to an object that is still in its construction phase. We
annotate with \texttt{@Raw} such variables.  As the invariant
described by the annotations is not yet established during the object
initialization, reading a field of an object annotated as
\texttt{@Raw} may return a null value (\texttt{@Nullable}) whatever is
the annotation on the field.  The example in
Fig.~\ref{fig:examples}(b) has been corrected in
Fig.~\ref{fig:examples}(c).



We refine those \texttt{@Raw} annotations by indicating the set of
classes for which a constructor has been executed.  Annotations
concerning fields declared in those classes can be considered as
already valid despite the fact the initialization of the object is not
yet completely finished.  The set of classes for which a constructor
has been executed can be represented by a single class as the
execution order of the constructors is constrained by the class
hierarchy.


\subsection{Sound Inference}
\label{sec:inference}

One of the key ideas behind non-null annotation inference is to track
field initialization in constructors and methods 
called from constructors.  At the end of constructors, all fields
defined in the current class which might not have been explicitly
initialized are annotated with \texttt{@Nullable} while all other
fields are annotated with the value they have been initialized with.

\begin{figure}[ht]
  \centering
  \begin{tabular}[t]{lrcl}
    Variables & $x$ & $\in$ & $\V$\\
    Fields & $f$ & $\in$ & $\F$\\
    Class names & $C$ & $\in$ & $\C$\\
    Labels & $l$ & $\in$ & $\PC$\\
    Method signatures & $m$ & $\in$ & $\M$\\
  \end{tabular}
  \begin{eqnarray*}
    E &::=& x \ |\ E.f \ \\
    I &::=&  x <- E \; | \; x.f <- E \; | \; x <- \bnew\ C(E,\ldots,E)\\
    & & | \; \bif\ (\star)\ l \;|\;  x.m(E,\ldots,E)  \; | \;  \bret
  \end{eqnarray*}
  \caption{Grammar of the Idealized Language}
  \label{fig:grammar}
\end{figure}

In~\cite{hubert08-1:nonnull_annotations_inference}, we chose to
formalize and prove our analysis on a language close to the bytecode
but without an operand stack.  This trade-off came from the fact that
we planed an implementation at the bytecode level but removing the
stack avoided to deal with alias information, which simplified the
presentation and the proofs.  Figure~\ref{fig:grammar} shows the
grammar of the idealized language.

To build our provably sound inference analysis we defined an
\emph{abstract domain} $\ab\State$ and a \emph{constraint based
  specification} of the analysis that constrain
$\ab{\text{S}}\in\ab\State$ depending on a program $P$ (denoted by
$\ab{\text{S}}|=P$).
To prove the soundness of the analysis we completed several tasks.

\begin{itemize}
\item We defined \emph{concrete domains} and a \emph{concrete
    semantics} for the language.
\item We explained \emph{how to interpret} $\ab\State$ in terms of
  concrete values with the relation $\sim\ \subseteq \ab\State \times
  \State$.
\item We defined the property $\safe(\semantics{P})$ which holds if
  all accessible states of the program $P$ are safe (but
  $\semantics{P}$ is not computable in general).
\item We defined the property $\ab\safe(\ab{\text{S}})$ which holds if
  $\ab{\text{S}}$ enforces $\safe(\semantics{P})$ (assuming
  $\ab{\text{S}}$ admits the states of $\semantics{P}$ as conservative
  interpretation, \emph{i.e.} $\forall
  \text{S}\in\semantics{P}.(\ab{\text{S}}\sim \text{S})$).
\item Finally, we proved the soundness, \emph{i.e.} we proved that if
  $\ab{\text{S}}$ is a solution of the constraint system for the
  program $P$ ($\ab{\text{S}}|=P$) and if $\ab\safe(\ab{\text{S}})$
  holds, then $\safe(\semantics{P})$ holds.  ($\ab\safe(\ab{\text{S}})
  \; \land \; \ab{\text{S}}|=P \; => \; \safe(\semantics{P})$)
\end{itemize}

In the following, we will first give some details about the concrete
and abstract domains and the relation between them.  Then, we will
show some examples of constraint rules of the analysis.

\subsubsection{Concrete and Abstract Domains}
\label{sec:domains}

\begin{figure}
  \centering
    \begin{eqnarray*}
      \Val &=& \Loc + \{ \NULL \} \\
      \Object &=& \F \rightharpoonup \Val \times \{ \isdef, \isundef\}\\
      \LocalVar &=& \V \to \Val + \{\bot\} \\
      \Heap &=& \Loc \pto \Object \times \C \times \wp(\C)\\
      \State &=& (\PC \times \LocalVar \times \Heap) + \Error
    \end{eqnarray*}
  \caption{Concrete Domains}
  \label{fig:concrete-domains}
\end{figure}
The concrete domains are the domains manipulated by the operational
semantics of the language.  Their definitions in
Fig.~\ref{fig:concrete-domains} are standard except for two particular
additions.  In order to reason about object and field initialization,
we instrumented the semantics and the domains so 1) each field of each
object has a flag which indicates if the field has been (explicitly or
implicitly) initialized, and 2) the set of classes for which a
constructor has been executed is attached to each object.  This
addition is used to prove the correctness of the refinement of the
\texttt{@Raw} annotation.

\begin{figure*}[!t]
  \centering
  \begin{displaymath}
    \begin{array}{cc}
      \begin{array}{r@{\ }c@{\ }l}
        \ab{\Val} & = &  \{\rawtop, \NotNull, \Null \} \\
        & & \cup\ \{ \raw(Y)\ |\ Y \in \Class \}\\
        \ab{\Def} &=& \{\Def, \UnDef\}\\
        \TVal &=& \F \pto \ab{\Def}\\
        \ab{\Heap} &=& \F \to \ab{\Val}\\
        \ab\LocalVar &=& \V \to \ab\Val\\
        \ab\Method &=& \M \to
        \struct{\mathsf{this}\in \TVal;\mathsf{args}\in{(\ab{\Val})}^{*};\mathsf{post}\in \TVal}\\
        \ab{\mathrm{State}} &=&  \ab\Method \times \ab{\Heap} \times \left(\M \times \PC \pto \TVal\right)\\
        && \times \left(\M \times \PC \pto \ab\LocalVar \right)
      \end{array}
      &
      \begin{array}[c]{c}
        \inference{v\in\Val}{\Null \sim_{h} v} \\[2ex]
        \inference{
          v\in\dom(h) & 
          \forall f\in\dom(h(v)), \IsDef(h(v)(f))}{\NotNull \sim_{h} v}\\[2ex]
        
        \inference{v\not=\NULL}{\rawtop \sim_{h} v}\\[2ex]
        \inference{
          v\in\dom(h) & 
          \forall f\in\bigcup_{A\preceq C}\fields(C), \IsDef(h(v)(f))}{\raw(A) \sim_{h} v}
      \end{array}
    \end{array}
  \end{displaymath}
  \caption{Abstract Domains and Selected Correctness Relations}
  \label{fig:abstract-domains}
\end{figure*}

Figure~\ref{fig:abstract-domains} presents the abstractions of the
concrete domains.
The $\ab\Val$ domain represents the annotation described in
Sect.~\ref{sec:nonnull-annotations}: references are either abstracted
with $\NotNull$, $\Null$ or $\Raw$.
Figure~\ref{fig:abstract-domains} also gives the correctness relations
for $\ab\Val$, which express how to interpret $\ab\Val$ in terms of
concrete values.  The first rule defines $\Null$ as $\top$,
\emph{i.e.}  $\Null$ abstracts any value.  The second rule defines
$\NotNull$ as an abstraction of references to objects that have all
their fields initialized.  The third rule defines $\rawtop$ as an
abstraction of any non-null value.  Finally, the forth rule defines
$\raw(A)$ as an abstraction of references to objects that have all
their fields defined in \emph{A} or super-classes of \emph{A}
initialized.

The $\ab\Def$ domain is used by $\TVal$ to represent the
initialization state of the fields of the current object
(\texttt{this} in Java) that are defined in the current class. At the
beginning of the object's lifetime, the abstraction of the current
object $\T\in\TVal$ associate to each field defined in the current
class the abstract value $\UnDef$.  The abstraction then evolves as
fields are initialized.  The reason for limiting the information to
fields defined in the current class (and not to all fields of
instances of the current class) is to keep the checker modular and
annotations easy to understand by a developer.
The abstraction used for the heap does not differentiate objects and
is flow-insensitive.  This is quite standard and corresponds to the
purpose of the analysis, \emph{i.e.} giving one annotation for each
field.
As the annotations must be easy to read, they are context-insensitive,
but, to achieve some precision, the analysis is inter-procedural and
method signatures are inferred from the join of the calling contexts
(as in 0-CFA~\cite{54007}).  A method signature includes the
initialization state of the fields of the current object
($\mathsf{this}$) and an abstract value for each parameter
(\textsf{args}).  A method signature also represents the result of the
method, \emph{i.e.} the fields that are initialized at the end of the
method ($\mathsf{post}$).  The analysis has been designed without
return values but adding them is straightforward.
Those domains are then combined to form the abstract state that
correspond to all reachable state of the program.  To be able to use
strong updates~\cite{chase90:pointer_analysis}, \emph{i.e.} to
precisely analyse assignments, the abstractions of the local variables
and the current object are flow-sensitive while, as discussed earlier,
the abstraction of the heap is flow-insensitive.

\subsubsection{Constraint based data flow analysis}
\label{sec:inferrence}
The analysis computes $\ab{\text{S}}\in\ab\State$ such that
$\ab{\text{S}} |= P$, \emph{i.e.} $\ab{\text{S}}$ over-approximates
all reachable states of the program $P$.  We have specified this
property as a constraint based data flow analysis.
Expressing the analysis in terms of constraints over lattices has the
immediate advantage that inference can be obtained from standard
iterative constraint solving techniques for static analyses.

\begin{figure*}
  \centering
$$
\begin{array}[c]{cl}
\inference
  {
    \forall \meth, \meth', \mathrm{overrides}(\meth',\meth) => \MV(\meth)[\mathsf{args}]\sqsubseteq \MV(\meth')[\mathsf{args}]\\
    \forall \meth, \meth', \mathrm{overrides}(\meth',\meth) => \top_{\class(\meth')} \sqsubseteq \MV(\meth')[\mathsf{this}]  \\
    \forall \meth, \meth', \mathrm{overrides}(\meth',\meth) => \top_{\class(\meth)} \sqsubseteq \MV(\meth)[\mathsf{post}]  \\
    \forall \meth,\  \MV(\meth)[\mathsf{this}]\sqsubseteq \T(\meth,0) \qquad
    \forall \meth,\  \MV(\meth)[\mathsf{args}]\sqsubseteq \LV(\meth,0)\\
    \top_{\class(\main)} \sqsubseteq \T(\main,0)    \qquad    \rawtop \sqsubseteq \LV(\main,0)(\this)\\
    \forall \meth,\ \forall i,\ \MV, \H, \T, \LV |= (\meth,i) ~:~P_\meth[i]}
  {\MV, \H, \T, \LV |= P} & (1)\\[3ex]
  \inference
  {\text{if }x = \this \land f\in\fields(\class(\meth))\text{ then } \T(\meth,i)[f\mapsto\Def] \text{ else }
    \T(\meth,i) \sqsubseteq \T(\meth,i+1)\\
   \LV(\meth,i) \sqsubseteq \LV(\meth,i+1) &
   \ab{\ia{e}} \sqsubseteq \H(f)}
  { \MV, \H, \T, \LV |= (\meth,i) ~:~x.f <- e}
&(2)\\[3ex]
  \inference
  {\T(\meth,i) \sqsubseteq \MV(\meth)[\mathsf{post}] \\
     \name(\meth)=\minit => 
       \forall f\in \fields(\class(\meth)).(\T(\meth,i)(f)=\UnDef =>
         \Null \sqsubseteq \H(f))
    }
  { \MV, \H, \T, \LV |= (\meth,i) ~:~\bret}
&(3)
\end{array}
$$
  \caption{Some Constraint Based Rules}
  \label{fig:rules}
\end{figure*}

To simplify the rules, we denote by $\MV, \H, \T, \LV$ a value of
$\ab\State$.  The main rule of the analysis is Rule~(1) given in
Fig.~\ref{fig:rules}.  The first two constraints of this rule state
that annotations on method arguments are contravariant, \emph{i.e.}
the lower in the hierarchy the less precise is the annotation.  It is
standard in object oriented languages as a virtual call to a method in
a top-level class may dynamically be resolved to a call in one of its
sub-classes.
%
The third constraint is conversely implied by the covariance of the
method post-condition (\textsf{post}).
The next two constraints link the method signatures with the
flow-sensitive information used by the intra-procedural part of the
analysis.
Finally, the last line constrains the state depending on all
instructions of the program using a judgment of the form $$\MV, \H,
\T, \LV |= (\meth,i) ~:~\mathit{instr}$$ for when the abstract state
$\MV, \H, \T, \LV$ is coherent with instruction $\mathit{instr}$ at
program point $(\meth,i)$. The two other rules are examples of such
judgements.~\footnote{The complete judgment set can be found
  in~\cite{hubert08-1:nonnull_annotations_inference}.  }

Rule~(2) corresponds to the instruction that sets the field $f$ of the
object stored in the local variable $x$ to the value of the expression
$e$.  The local variables are unchanged by such an instruction and the
abstraction of the expression $e$ is computed ($\ab{\ia{e}}$) and
added to the possible values of $f$.  Then, depending on whether $x$
is \texttt{this} or not, the flow-sensitive information $\T$ may be
updated to reflect the fact that a field of the current object has
been initialized.

The judgment for the return instruction is given in Rule~(3).  If the
instruction is found in a constructor then the abstraction of the null
constant ($\Null$) is added to all fields that are not labeled as
initialized at the end of the constructor.

\section{Towards a JVML Analysis}
\label{sec:bytecode-language}
In~\cite{hubert08-1:nonnull_annotations_inference}, we presented a
first prototype of the analysis and some features that were mandatory
to target the Java bytecode language (JVML).  The prototype already
had a simple must-alias analysis to track the references to
\texttt{this} in the stack and some other conservative features.  This
section proposes three modifications of the analysis on the JVML.

The JVML is a stack language and includes some instructions to test
variables for the \texttt{null} constant, such as \texttt{ifnull~n}
which pops the top of the stack and jumps \texttt{n} bytes of
instructions if the popped reference is null.
From such an instruction, the analysis infers that, when the test
fails, the popped value is non-null but this information is useless to
the analysis as this value cannot be reread.  For the information to
be exploitable, the analysis must know to which local variable the
popped value was equal
to
.  To infer equalities between stack and local variables we have
implemented a must-alias analysis.
%
%
For example, in Fig.~\ref{fig:tests}, assuming \texttt{x} is annotated
as \texttt{@Nullable} at the beginning, this allows to infer that the
second load loads a non-null (\texttt{@Raw}) value.
\begin{figure}
  \centering
  \begin{tabular}{l@{\hspace*{5ex}}l}
\tt
load x & $x \mapsto \Null$\\
\tt
ifnull n & $x \mapsto \NotNull$\\
\tt
load x & $x \mapsto \NotNull$\\
\multicolumn{2}{c}{...}\\
\end{tabular}
  \caption{Recovering Information from Tests}
\label{fig:tests}
\end{figure}

Assume we have two functions $\alpha \in 2^\Val -> \ab\Val$ and
$\gamma \in \ab\Val -> 2^\Val$ which computes the abstraction of a set
of concrete values and the concretization of an abstract value,
respectively.
In Fig.~\ref{fig:tests}, assume that \texttt{x} is either non-null and
fully initialized or null at the beginning of the example.  The
analysis abstracts such a value by
$$\NotNull \sqcup \alpha(\{\mathtt{null}\}) = \Null.$$
The test allows to recover some information but, as $\Null$ also
abstract raw references, the most that can be recovered is
$$\alpha(\gamma(\Null) \setminus \{\mathtt{null}\}) = \rawtop.$$
Such configurations often occur in real programs as implicitly
initialized fields are always annotated with $\Null$, despite they may
never contain any raw value.  To solve this, we add $\NullInit$, a new
abstract value that abstract values that may not point to raw
objects. We have
$$
\begin{array}{c}
\NotNull \sqsubset \NullInit \sqsubset \Null\\
\alpha(\gamma(\NullInit) \setminus \{\mathtt{null}\}) = \NotNull~.
\end{array}
$$
It allows to annotate more references as \texttt{@NonNull} and
therefore to gain in precision as field annotations can then be
trusted.  It also allows a more direct gain in precision.  A variable
annotated as \texttt{@NullableInit} may not point to an object that is
not fully initialized, so field annotations can also be trusted when
reading fields of variables annotated with \texttt{@NullableInit}.

The JVML includes the \texttt{instanceof C} instruction which pushes 1
on the stack if the top of the stack is non-null and is an instance of
the class \texttt{C}, otherwise it pushes 0.  A conditional jump
generally occurs few instructions after.
Recovering information from such an instruction is not trivial: both
the \texttt{instanceof} instruction and the jump are needed, they may
be separated by some other instructions, and they interact in the
concrete semantics through integer values.
To be able to use this information, we have defined another analysis
which computes an abstraction of the stack such that, for each stack
variable, it contains an under-approximation of the set of local
variable that must be non-null if the corresponding stack variable is
equal to 1.

\label{sec:empirical-results}
\begin{table*}
  \small
  \centering
  \begin{tabular}{|l|l|r|r|r|r|r|r|r|r|}
    \cline{3-10}
    \multicolumn{2}{l|}{} & \multicolumn{2}{c|}{Fields} & \multicolumn{2}{c|}{Parameters} & \multicolumn{2}{c|}{Return}& \multicolumn{2}{c|}{Total}\\
    \cline{3-10}
    \multicolumn{1}{l}{Project} & Version & \multicolumn{1}{c|}{\#} & \multicolumn{1}{c|}{Non-null(\%)} & \multicolumn{1}{c|}{\#} & \multicolumn{1}{c|}{Non-null(\%)} & \multicolumn{1}{c|}{\#} & \multicolumn{1}{c|}{Non-null(\%)} & \multicolumn{1}{c|}{\#} & \multicolumn{1}{c|}{Non-null(\%)}\\\hline

    \multirow{2}*{Jess}
    &\fmoods{} & 319 & 48.6 & 1663 & 29.2 & 789 & 44.7 & 2771 & 35.8\\
    & \optimized{}  & 319 & 55.8 & 1660 & 50.1 & 788 & 51.1 & 2767 & 51.0\\\hline
    \multirow{2}*{Soot}
    & \fmoods{}  & 3457 & 54.2 & 9793 & 44.5 & 4177 & 57.3 & 17427 & 49.5\\
    & \optimized{}  & 3456 & 58.0 & 9793 & 54.5 & 4177 & 63.4 & 17426 & 57.3\\\hline
    \multirow{2}*{ESC/Java}
    & \fmoods{}  & 746 & 49.5 & 3075 & 28.0 & 1155 & 50.6 & 4976 & 36.5\\
    & \optimized{}  & 744 & 52.6 & 3067 & 42.6 & 1152 & 58.6 & 4963 & 47.8\\\hline
    \multirow{2}*{Julia}
    & \fmoods{}  & 396 & 39.4 & 1481 & 36.1 & 842 & 52.1 & 2719 & 41.5\\
    & \optimized{}  & 396 & 47.2 & 1481 & 48.3 & 842 & 63.4 & 2719 & 52.8\\\hline
    \multirow{2}*{JDTCore}
    & \fmoods{}  & 1018 & 45.0 & 3526 & 24.7 & 916 & 38.9 & 5460 & 30.9\\
    & \optimized{}  & 1018 & 47.0 & 3525 & 42.1 & 916 & 46.9 & 5459 & 43.8\\\hline
    \multirow{2}*{JavaCC}
    & \fmoods{}  & 116 & 50.0 & 292 & 33.2 & 85 & 77.6 & 493 & 44.8\\
    & \optimized{}  & 116 & 50.9 & 292 & 43.5 & 85 & 82.4 & 493 & 51.9\\\hline
    \multirow{2}*{Others}
    & \fmoods{}  & 783 & 50.2 & 2244 & 43.5 & 717 & 43.2 & 3744 & 44.8\\
    & \optimized{}  & 783 & 56.1 & 2244 & 56.0 & 717 & 50.6 & 3744 & 55.0\\\hline
    \hline
    \multirow{2}*{Total}
    & \fmoods{}  & 8062 & 48.9 & 26662 & 36.3 & 10592 & 48.8 & 45316 & 41.5\\
    & \optimized{}  & 8059 & 52.7 & 26650 & 48.7 & 10588 & 55.1 & 45297 & 50.9\\\hline
  \end{tabular}
  \caption{Annotation Results}
  \label{tab:results-annot}
\end{table*}

Finally, when a possibly null value is dereferenced, if the control
flow reaches the next instruction it means that, at this point, the
reference is non-null.  Therefore, it is possible to refine all
instructions that dereference variables so those variables are
inferred as non-null on outgoing non-exceptional edges of the control
flow graph.

\section{Implementation}
\label{sec:details}
The global inferencer is a whole program analysis composed of three
analysis: the alias analysis, the analysis of \texttt{instanceof}
instructions and the main non-null analysis described in this paper.
The non-null analysis uses the results of both the alias and the
\texttt{instanceof} analyses and the \texttt{instanceof} analysis uses
the results of the alias analysis.  Those communications between the
analyses impose the simple scheduling of running first the alias
analysis, then the \texttt{instanceof} analysis and in the end the
non-null analysis.  The three analyses have been implemented in a
similar standard fashion.  First we iterate over all instructions of
the program to build transfer functions that take as argument a part
of the abstract state and that return the parts of the abstract states
that have been modified by the instructions. We store the functions in
a hash-map with their dependencies as keys and we apply a work list
algorithm to compute the fixpoint.
The result of each analysis is the fixpoint computed.

The global result contains, for each program point, three
abstractions, one of which, the non-null abstraction, containing
already a lot of information.  Such an analysis cannot be implemented
without taking care of memory consumption.
We have done the implementation in OCaml~\cite{ocaml} so we have been
able to use JavaLib~\cite{javalib} ---~a \texttt{.class} file parser
we maintain.
%
We have put a lot of coding effort in reducing the memory consumption.
We have implemented $\ab\LocalVar$, $\TVal$ and $\ab\Heap$ as balanced
binary trees, which, as well as being efficient, have easily allowed
us not to store bottom values ($\NotNull$ and $\Def$).  This is
specially important as non-reference type are coded as bottom and most
variables are non-null.
We use sharing extensively and functional programming has greatly
helped us herein.  \emph{E.g.} the stack is implemented as a list and,
between two instructions, the part of the stack that is unchanged by
the instruction is shared in memory and, to some extent, the same
applies to maps.  Using sharing has also improved efficiency as it has
then been possible to use physical equality tests instead of
structural equality tests in some places.
The result of the alias and \texttt{instanceof} analysis are compacted
to remove the information for the instructions we know the results
will not be used.  \emph{E.g.} for the result of the
\texttt{instanceof} analysis, only the abstractions of stacks at
conditional jumps ($\sim$ 5\% of the instructions) are kept.

\section{Empirical Results}

\begin{table*}
  \small
  \centering
  \begin{tabular}{|l|l|r|r|r|r|r|r|r|r|r|r|}
    \cline{3-12}
    \multicolumn{2}{l|}{} & \multicolumn{2}{c|}{Field Read} & \multicolumn{2}{c|}{Field Write} & \multicolumn{2}{c|}{Method Call} & \multicolumn{2}{c|}{Array Operations}& \multicolumn{2}{c|}{Total}\\
    \cline{3-12}
    \multicolumn{1}{l}{Project} & Version & \multicolumn{1}{c|}{\#} & \multicolumn{1}{c|}{Safe(\%)} & \multicolumn{1}{c|}{\#} & \multicolumn{1}{c|}{Safe(\%)} & \multicolumn{1}{c|}{\#} & \multicolumn{1}{c|}{Safe(\%)} & \multicolumn{1}{c|}{\#} & \multicolumn{1}{c|}{Safe(\%)} & \multicolumn{1}{c|}{\#} & \multicolumn{1}{c|}{Safe(\%)}\\
    \hline

\hline\multirow{2}*{Jess}
& \fmoods{} & 2725 & 85.4 & 919 & 97.7 & 10358 & 63.0 & 729 & 73.9 & 14731 & 69.8\\
& \optimized{} & 2721 & 97.4 & 917 & 98.0 & 10330 & 76.7 & 727 & 83.2 & 14695 & 82.2\\
\hline\multirow{2}*{Soot}
& \fmoods{} & 25323 & 76.5 & 7038 & 92.7 & 89161 & 73.4 & 3092 & 57.3 & 124614 & 74.7\\
& \optimized{} & 25319 & 80.7 & 7037 & 95.0 & 89138 & 82.2 & 3092 & 69.8 & 124586 & 82.3\\
\hline\multirow{2}*{JDTCore}
& \fmoods{} & 29200 & 75.2 & 7569 & 85.8 & 25205 & 50.2 & 10769 & 21.6 & 72743 & 59.7\\
& \optimized{} & 29196 & 88.3 & 7568 & 92.3 & 25197 & 72.6 & 10765 & 45.9 & 72726 & 77.0\\
\hline\multirow{2}*{JavaCC}
& \fmoods{} & 5733 & 81.7 & 1096 & 92.7 & 11772 & 66.0 & 2129 & 50.4 & 20730 & 70.2\\
& \optimized{} & 5733 & 92.6 & 1096 & 95.8 & 11772 & 71.8 & 2129 & 65.9 & 20730 & 78.2\\
\hline\multirow{2}*{ESC/Java}
& \fmoods{} & 10799 & 50.3 & 2570 & 90.9 & 19129 & 67.0 & 1441 & 54.3 & 33939 & 63.0\\
& \optimized{} & 10787 & 90.3 & 2568 & 98.4 & 19077 & 78.9 & 1441 & 79.9 & 33873 & 84.1\\
\hline\multirow{2}*{Julia}
& \fmoods{} & 4474 & 75.6 & 1065 & 90.0 & 15836 & 72.2 & 987 & 39.9 & 22362 & 72.3\\
& \optimized{} & 4474 & 82.9 & 1065 & 95.0 & 15835 & 85.2 & 987 & 55.0 & 22361 & 83.9\\
\hline\multirow{2}*{Others}
& \fmoods{} & 12341 & 72.8 & 3295 & 90.5 & 19189 & 66.4 & 3182 & 32.2 & 38007 & 67.7\\
& \optimized{} & 12341 & 79.3 & 3295 & 93.4 & 19182 & 76.4 & 3182 & 62.7 & 38000 & 77.7\\
\hline\hline\multirow{2}*{Total}
& \fmoods{} & 104496 & 76.2 & 26882 & 91.1 & 230574 & 67.5 & 25864 & 34.9 & 387816 & 69.3\\
& \optimized{} & 104468 & 87.2 & 26874 & 95.0 & 230447 & 78.3 & 25858 & 56.5 & 387647 & 80.4\\
\hline
   \end{tabular}
  \caption{Dereferencing Results}
  \label{tab:results-deref}
\end{table*}

The benchmarks includes production applications such as Soot
2.2.4~\cite{soot}, the JDT Core Compiler of Eclipse
3.3.3~\cite{eclipse}, Julia 1.4~\cite{julia}, ESC/Java
2.0b4~\cite{escjava2} and Jess 7.1p1~\cite{jess}.
It also includes some other smaller applications such as
JavaCC~\cite{javacc}, Jasmin~\cite{jvm}, Tight\-VNC
Viewer~\cite{tightvnc} and the 10 programs constituting the SPEC JVM98
benchmarks~\cite{spec98}.
The implementation used for those benchmarks is NIT 0.4, coded in
OCaml 3.10.2~\cite{ocaml}, and uses the JavaLib 1.7~\cite{javalib}
library.  We performed the whole-program analysis with the Java
Runtime Environment of GNU gcj 3.4.6~\cite{gnugcj} on a MacBook Pro
with a 2.4~GHz Intel Core 2 Duo processor with 4~GB of RAM.

In our implementation, we have added switches to enable the
modifications we have proposed in this article that may interact with
the precision.  The results obtained with all the modifications
enabled are labeled as \optimized{} and the version without
modifications \fmoods{}.
Table~\ref{tab:results-annot} gives the number of annotations and the
percentage of those which are non-null (\texttt{@Raw} or
\texttt{@NonNull}) for field annotations, method parameters (except
\texttt{this}) and method results.  Notice the two inferencers do not
give the same number of annotations: the more precise the analysis is
the more dead code is removed.  The improvement between the two
analyses is substantial: while \fmoods{} inferred 41.5\% of non-null
annotations, \optimized{} inferred 50.9\% of non-null annotations,
\emph{i.e.}  the improvement brought by \optimized{} over \fmoods{}
represents more than 9 points.  Chalin and James experimentally
confirmed~\cite{DBLP:conf/ecoop/ChalinJ07} that at least two thirds of
annotations in Java programs in general are supposed to be non-null.
With regard to their experiment, 50.9\% is already a significant
fraction.

The purpose of annotations is to reduce the number of potential
exceptions, so we instrumented our inferencer to count the number of
dereferences that were proved safe with the inferred annotations.
Table~\ref{tab:results-deref} shows the percentage of safe
dereferences over the total number of dereferences for field reads,
field writes, method calls and array operations (load, store and
length).  Note how \fmoods{} implementation was able to prove 2/3 of
dereferences safe and how \optimized{} reduced the number of unsafe
dereferences by a 1/3, globally proving more than 80\% of dereferences
in the studied benchmarks safe.

Finally, Tab.~\ref{tab:results-perf} gives the memory and time
consumption for the most expensive benchmarks and the sum and the
maximum memory and time consumption for the other benchmarks for the
\optimized{} inferencer.  Assuming enough processors and memory, the
analyses can be run in parallel so the resources needed correspond to
the sum of the memory consumption but the maximum of time consumption,
while if the analyses are run sequentially, the resources needed are
the maximum memory consumption but the sum of time consumption.
%
%
%
The
worst case is the analysis of Jess (804,111 bytecode instructions
including libraries and excluding dead code), which needs 887~MB and
144~s.
%
%
Our results indicate that the implementation scales.

\begin{table}
  \centering
  \small
  \begin{tabular}{|l|l|r|r|}
    \cline{3-4}
    \multicolumn{2}{l|}{} & \multicolumn{1}{c|}{Space (MB)} & \multicolumn{1}{c|}{Time (s)}\\
    \hline
    \multicolumn{2}{|l|}{Jess} &887&144\\\hline
    \multicolumn{2}{|l|}{Soot} &634&122\\\hline
    \multicolumn{2}{|l|}{ESC/Java} &421&51\\\hline
    \multicolumn{2}{|l|}{Julia} &363&44\\\hline
    \multicolumn{2}{|l|}{JDTCore} &331&36\\\hline
    \multicolumn{2}{|l|}{JavaCC} &310&34\\\hline
    \multirow{2}*{Others}&sum&1642&159\\\cline{2-4}
    &max&253&30\\\hline
  \end{tabular}
  \caption{Time and Space Consumption}
  \label{tab:results-perf}  
\end{table}



\section{Related work}
\label{sec:related-work}

\leino{} proposed the non-null type system~\cite{FahndrichL03} on
which this analysis is transitively based.  Papi \emph{et al.}
propose a framework~\cite{PapiACPE2008} for Java source code
annotations and, as an example, provide a checker for non-null
annotations based on~\cite{FahndrichL03}.  Ekman \emph{et al.}
propose an plug-in~\cite{JastAddNonNull} for JastAdd~\cite{EkmanPhD}
to infer and check the same non-null annotations.  As their work
predates the improvements we proposed
in~\cite{hubert08-1:nonnull_annotations_inference} and as we further
improved the analysis in this paper, our analysis is strictly more
precise.  F\"ahndrich and Xia~\cite{fahndrich07:delayed_types} propose
another analysis to deal with object initialization which can deal
with circular data structure but the generality of their framework
prevents the analysis from being as precise as our on examples without
circular data structures.  Furthermore, their analysis can only infer
a part of the annotations needed.

To infer type annotations, Houdini~\cite{flanagan01houdini} generates
a set of possible annotations (non-null annotations among others) for
a given program and uses ESC/Java~\cite{escjava} to refute false
assumptions.
CANAPA~\cite{CieleckiFJJ06} lowers the burden of annotating a program
by propagating some non-null annotations.  It also relies on ESC/Java
to infer where annotations are needed.  Those two annotation
assistants have a simplified handling of objects under construction
and rely on ESC/Java~\cite{escjava}, which is neither sound nor
complete.
%
Some other work has focused on local type inference, \emph{i.e.}
inferring nullness properties for small blocks of code like methods.
One example hereof is the work of Male \emph{et
  al.}~\cite{male08:_java_bytec_verif_for_types}.
%
Spoto very recently proposed another nullness inference
analysis~\cite{spoto08:nullness_analysis} with a different domain
that expresses logical relations between nullness of variables.  He
compared his implementation with an old version of our tool NIT which
did not include
improvements on precision and performance we made.  We herein included
in the benchmarks some of the programs used
in~\cite{spoto08:nullness_analysis} and we can notice that, despite
the context sensitivity of the analysis
in~\cite{spoto08:nullness_analysis}, both analyses have very close
practical results.

FindBugs~\cite{hovemeyer06:_tunining_static_analysis_to_fingbugs,hovemeyer07:_findin_more_null_pointer_bugs}
and Jlint~\cite{artho-vmcai04} use static analyses to find null
pointer bugs.  To keep the false positive and false negative rates low
they are neither sound nor complete.

\section{Conclusion}
\label{sec:conclusion}
We have proposed an improved version of our provably sound inference
analysis along with an efficient implementation of both analyses.
Despite being a whole-program analysis, it is possible to infer
annotations for production programs within minutes.  The precision of
the analysis is not yet sufficient to certify existing code without
handwork afterwards, but it is still of interest for code
documentation, for reverse engineering and for improving the precision
of control flow graphs, which is useful to native code compilers and
other program verifications and static analyses.

While we still plan to further improve the inference analysis, we
believe the need for a certified checker at the bytecode level is
bigger than ever and hence this is in our priorities.

\appendix
\bibliographystyle{acm}
\bibliography{biblio}

\end{document}